\documentclass[oribibl]{llncs}

\usepackage{amssymb}
\usepackage{graphicx}
\usepackage{times}
\usepackage{helvet}
\usepackage{courier}

\usepackage{url}
\usepackage{amsmath}
\usepackage{amssymb}
\usepackage{color}
\usepackage{epsf}
\usepackage{color}
\usepackage{graphicx}
\usepackage{epsfig}
\usepackage{fancybox}
\usepackage{multicol}
\usepackage{hhline}
\usepackage{xspace,epic,eepic,graphicx}
\usepackage{latexsym}
\usepackage[all]{xy}
\DeclareMathAlphabet{\mathpzc}{OT1}{pzc}{m}{it}

\usepackage{floatflt}
\usepackage{wrapfig}

\def\be{\begin{equation}}
\def\ee{\end{equation}}
\def\bea{\begin{eqnarray}}
\def\eea{\end{eqnarray}}

\def\ba{\begin{align}}
\def\ea{\end{align}}
\def\bes{\begin{split}}
\def\es{\end{split}}

\def\mymathhyphen{{\hbox{-}}}

\AtBeginDocument{%
  \begingroup\lccode`~=`_%
  \lowercase{\endgroup\let~}_%
  \catcode`_=12
}

\mathchardef\myhyphen="2D

\newcommand{\boxtheorem}{\hfill $\blacksquare$\vspace{1mm}}
\newcommand{\ignore}[1]{}
\newcommand{\nit}[1]{{\it #1}}

\DeclareMathAlphabet{\mathpzc}{OT1}{pzc}{m}{it}
\newcommand{\match}{{\mathpzc m}}

\newcommand{\eat}[1]{}
\newcommand{\mc}[1]{\mathcal{ #1}}

\newcommand{\red}[1]{\textcolor{red}{#1}}

\newcommand{\comlb}[1]{{\vspace{2mm}\noindent \bf \red{COMM(LEO):}}~ #1 \hfill {\bf
    END.}\\}

\newcommand{\comzb}[1]{{\vspace{2mm}\noindent \bf \red{\newline COMM(Zeinab):}}~  #1 \hfill {\bf
    END.}\newline \\}

\newcommand{\ent}[1]{{\sf \small #1}}

\newcommand{\mf}[1]{\mathbf{#1}}

\begin{document}
\pagestyle{plain}
\thispagestyle{empty}

\title{ {\em ERBlox}: \ Combining  Matching Dependencies with Machine Learning for Entity Resolution}

\author{{\bf Zeinab Bahmani}\inst{1}, \ {\bf Leopoldo Bertossi}\inst{1} \  and \ {\bf Nikolaos Vasiloglou}\inst{2}}

\institute{Carleton University, \ School of Computer Science, \ Ottawa, \ Canada
\ignore{\email{\{mmilani,bertossi\}@scs.carleton.ca}} \and LogicBlox Inc., \ Atlanta, \ GA 30309, \ USA}


%


%
%
\maketitle


\begin{abstract}
Entity resolution (ER), an important and common data cleaning problem, is about
detecting data duplicate representations for the same external entities,
and merging them into single  representations. Relatively recently, declarative rules called
{\em matching dependencies} (MDs) have been proposed for specifying similarity conditions
under which attribute values in database records are merged.  In this work we show the process and the benefits of
integrating three components of ER:  (a) Classifiers for duplicate/non-duplicate record pairs built
using machine learning (ML) techniques,  (b) MDs for supporting both the blocking phase of ML and
the merge itself; and (c) The use of the declarative language {\em LogiQL} -an extended form of Datalog
supported by the {\em LogicBlox} platform- for data processing, and the specification and
enforcement of MDs.
\end{abstract}

\vspace{-7mm}\hspace*{0.5cm}{\scriptsize {\bf Keywords:} \ Entity resolution,  matching dependencies, support-vector machines, classification, Datalog}

\section{Introduction}\label{intro}

\vspace{-2mm}
Entity resolution (ER) is a common and difficult problem in data cleaning that has to do with handling
unintended multiple representations in a database of the same external objects. Multiple representations
lead to uncertainty in data and the problem of managing it. Cleaning the database
reduces uncertainty. In more precise terms, ER
is about the identification  and fusion of  database records (think of rows or tuples in tables) that represent the same real-world
entity \cite{naumannACMCS,elmargamid}. As a consequence, ER usually goes through two main consecutive phases: (a) detecting duplicates, and (b) merging them into single representations.

\ignore{\paragraph{\bf Duplicate detection.}} For duplicate detection, one must first analyze multiple pairs of records, comparing the two records in them, and discriminating between: {\em pairs of duplicate records} and {\em pairs of non-duplicate records}. This classification problem is
approached with machine learning (ML)  methods,  to learn  from previously known or already made classifications (a training set for supervised learning), building a {\em classification model} (a classifier) for deciding about other record pairs \cite{Christen2007,elmargamid}.

In principle, in ER every two records (forming a pair) have to be compared, and then classified.
Most of the work on applying ML to ER work at the record level \cite{Rastgoi11, Christen2007, Christen2008}, and only some of the attributes, or their features, i.e. numerical values associated to them, may be involved in duplicate detection. The choice of relevant sets of attributes and features is application
dependent.

ER may be a task of quadratic complexity since it requires comparing every two records.  To reduce the large number two-record comparisons, {\em blocking techniques}
are used \cite{Baxter03,Herzog07,Garcia-Molina09}. Commonly, a single record attribute, or a combination of attributes, the so-called {\em blocking key}, is used to split the database records into blocks. Next, under the assumption that any two records in different blocks are unlikely to be duplicates, only every two records in a same block are compared for duplicate detection. \ignore{For example, we might block a set of employee records according to the city. We then only need to compare the employees with the same city to detect duplicate employees. Employee pairs that are in different blocks, with different cities, are considered as non-duplicate employee pairs.}

Although blocking will discard many record pairs that are obvious non-duplicates, some true duplicate pairs might  be missed (by putting them in different blocks),
due to errors or typographical
variations in attribute values. More interestingly, similarity between blocking keys alone may fail to capture the
relationships that naturally hold in the data and could  be used for blocking. Thus, entity blocking  based only on blocking key similarities may cause low recall.
This is a major drawback of traditional blocking techniques.

In this work we consider  different and coexisting entities. For each of them, there is a collection of records. Records for different entities may be related
via attributes in common or referential constraints. Blocking can be performed on each of the participating entities, and the way records for an entity are placed in blocks
may influence the way the records for another entity are assigned to blocks. This is called ``collective blocking". Semantic information, in addition to that provided
by blocking keys for single entities, can be used to
state relationships between different entities and their corresponding similarity criteria. So, blocking decision making forms a  collective and intertwined process involving several entities. In the end,
the records for each individual entity will be placed in blocks associated to that entity.

\ignore{\red{Collective blocking is a natural extension
of the notion of blocking in which one wants to block different types
of real-world entities in a set of records at the same time, in an intertwined process. In other words, in collective blocking, the semantic knowledge is used to make all the blocking decisions collectively, in addition to blocking keys which are used to block a single entity type. In general, the output of collective blocking is sets of blocks of the input records (by entity type).} }

\begin{example} \label{ex:IntroBlock} Consider two entities, \ent{Author} and \ent{Paper}. For each of them, there is a set of
records (for all practical purposes, think of database tuples in a single table). For \ent{Author} we have records of the form
$\mathbf{a} = \langle \nit{name}, \ldots, \nit{affiliation}, \ldots,$ $\nit{paper~title}, \ldots\rangle$, with \{\nit{name}, \nit{affiliation}\}  the blocking key; and for \ent{Paper}, records of the form
$\mathbf{p} = \langle \nit{title}, \ldots, \nit{author~name}, \ldots\rangle$, with \nit{title} the blocking key.
We want to group {\sf Author} and {\sf Paper} records at the same time, in an entwined process. We block together two \ent{Author} entities on the basis of the similarities of authors' names and  affiliations.

Assume that \ent{Author} entities $\mathbf{a}_1, \mathbf{a}_2$ have similar  names, but their affiliations are not. So, the two records would not be put in the same block.
However, $\mathbf{a}_1, \mathbf{a}_2$ are authors of  papers (in {\sf Paper} records) $\mf{p_1}, \mf{p_2}$, resp., which have been put in the same block (of papers) on the basis of similarities of paper titles\ignore{and papers keywords ({\em multi-attribute blocking keys})}. In this case, additional {\em semantic knowledge} might specify that if two papers are in the same block, then corresponding {\sf Author} records that have similar author names should be put in the same block too. Then, $\mathbf{a}_{\!1}$ and  $\mathbf{a}_2$ would end up in the same block.

In this example, we are blocking \ent{Author} and \ent{Paper} entities, separately, but collectively and in interaction.
 \boxtheorem
\end{example}
 Collective blocking is based on blocking keys and {\em the enforcement} of semantic information about the {\em relational closeness} of entities \ent{Author}
and \ent{Paper}, which is captured by a set of {\em matching dependencies} (MDs). So, we propose ``MD-based collective blocking" (more on MDs right below).

After records are divided in blocks, the proper duplicate detection process starts, and is carried out by comparing every two records
in a block, and classifying the pair as ``duplicates" or ``non-duplicates" using the trained ML model at hand. In the end,  records in duplicate pairs are
 considered to represent the same external entity, and have to be {\em merged} into a single representation,
i.e. into a single record. This second phase is also application dependent. MDs were originally proposed to support this task.

Matching dependencies are declarative logical rules that tell us under what conditions of similarity between attribute values, any two records must have certain attribute values merged, i.e. made identical
\cite{Fan08,FanJLM09}. For example, the MD \vspace{-2mm}
\begin{equation}\nit{Dept}_{\!B}[\nit{dept}] \approx \nit{Dept}_{\!B}[\nit{dept}] \ \to \ \nit{Dept}_{\!B}[\nit{city}] \doteq \nit{Dept}_{\!B}[\nit{city}] \label{eq:md}
\end{equation}

\vspace{-3mm}\noindent
tells us that for any two records for entity (or relation or table) $\nit{Dept}_{\!B}$ that have similar values for attribute $\nit{dept}$ attribute,  their values
for attribute $\nit{city}$ should be matched, i.e. made the same.

MDs as introduced in \cite{FanJLM09} do not specify how to merge values. In \cite{icdt11,Bertossi12}, MDs were extended with {\em matching functions} (MFs). For a  data domain, an MF
specifies how to assign a value in common to two values. We adopt MDs with MFs in this work. In the end, the enforcement of MDs with MFs should produce a duplicate-free instance
(cf. Section \ref{pre} for more details).

MDs have to be specified in a declarative manner, and at some point enforced, by producing changes on the data. For this purpose, we use the {\em LogicBlox} platform, a data management system developed by the LogicBlox\footnote{ www.logicblox.com} company, that is centered around its declarative language, {\em LogiQL}. {\em LogiQL} supports relational data management and, among several other features  \cite{Aref15},  an extended form of Datalog with stratified negation \cite{ceri90}. This language is expressive enough for the kind of MDs considered in this work.\footnote{For arbitrary sets of MDs, we need higher expressive power \cite{Bertossi12}, such as that provided by answer
set programming \cite{Bahmani12}.}

In this paper, we describe our {\em ERBlox} system. It is built on top of the {\em LogicBlox} platform, and implements entity resolution (ER) applying to {\em LogiQL},  ML techniques,   and the specification and enforcement
of MDs. 
More specifically, {\em ERBlox} has three main components: \ (a)
 MD-based collective blocking, \ (b) ML-based duplicate detection, and \
(c) MD-based merging. \
The sets of MDs are fixed and different for the first and last components. In both cases, the set of MDs are {\em interaction-free} \cite{Bertossi12}, which results, for each entity, in the unique set of blocks, and eventually into a single,
duplicate-free instance \cite{Bertossi12}.  We use {\em LogicQL} to declaratively implement the two MD-based
components of {\em ERBlox}.

The blocking phase uses  MDs to specify the blocking strategy. They express conditions in terms of blocking key similarities and also relational closeness (the semantic knowledge) to assign two records to a same block (by making the block identifiers identical).
Then, under MD-based collective blocking different records of possibly several related entities are simultaneously assigned to blocks through
the enforcement of MDs (cf. Section \ref{MDBlocking} for details).

On the ML side, the problem is about detecting pairs of duplicate records. The ML algorithm is trained using record-pairs known to be duplicates or non-duplicates. We independently used three established classification algorithms: {\em support vector machines} (SVMs) \cite{Vapnik98}, {\em k-nearest neighbor} (K-NN) \cite{Cover67}, and {\em non-parametric Bayes classifier} (NBC) \cite{Baudat00}. We used the Ismion\footnote{http://www.ismion.com} implementations of them
due to the in-house expertise at LogicBlox. Since the emphasis of this work is on the use of {\em LogiQL} and MDs, we will refer only to our use of SVMs.


We experimented with our {\em ERBlox} system using as dataset  a snapshot of Microsoft Academic Search (MAS)\footnote{http://academic.research.microsoft.com. For comparison, we also tested our system with data from DBLP and Cora.} (as of January 2013) including $250$K authors and $2.5$M papers. It contains a training set.
The experimental results show that our system improves ER accuracy over traditional blocking techniques \cite{Fellegi69}, which we will call {\em standard} blocking, where just blocking-key similarities are used. Actually, MD-based collective blocking leads to higher precision and recall on the given datasets.

This paper is structured as follows. Section \ref{pre} introduces background on matching dependencies and their semantics, and SVMs. A general overview of the {\em ERBlox} system is presented in Section \ref{MLMDFramework}. The specific  components of {\em ERBlox} are discussed in Sections \ref{sec:init}, \ref{MDBlocking}, and \ref{Merging}. Experimental
 results are shown in Section \ref{evaluation}. Section \ref{conclude} presents  conclusions.

\vspace{-3mm}
\section{Preliminaries}\label{pre}

\vspace{-2mm}
\subsection{Matching dependencies}\label{sec:mds}

\vspace{-2mm}

We consider an application-dependent relational schema $\mc{R}$, with a data domain $U$.
For an attribute $A$, $\nit{Dom}_{\!A}$ is its finite domain. We assume predicates
do not share attributes, but different attributes may share a domain.
An instance $D$ for $\mc{R}$ is a finite set of ground atoms of the form $R(c_1,\ldots, c_n)$, with $R \in \mc{R}$,
$c_i \in U$.

We assume that each entity is represented by a relational predicate, and its tuples or rows
in its extension correspond to records for the entity. As in \cite{Bertossi12}, we assume records
have unique, fixed, global identifiers, {\em rids}, which are positive integers. This allows us to  trace changes of attribute values in records.
Record ids are placed in an extra attribute for $R \in \mc{R}$ that acts as a key. Then, records take the form $R(r,\bar{r})$, with $r$ the rid, and $\bar{r} =(c_1,\ldots, c_n)$.
  Sometimes we leave rids implicit, and sometimes we use them  to denote whole records: if $r$ is a record identifier in instance $D$, $\bar{r}$ denotes the record in $D$ identified by $r$.
Similarly, if $\cal{A}$ is a sublist of the attributes of predicate $R$, then $r[\mc{A}]$ denotes the restriction of $\bar{r}$ to $\cal{A}$.

MDs  are formulas of the form: \ $R_1[\bar{X}_1] \approx R_2[\bar{X}_2] \ \rightarrow \ R_1[\bar{Y}_1] \doteq R_2[\bar{Y}_2]$ \  \cite{Fan08,FanJLM09}. \
 Here, $R_1, R_2 \in \mc{R}$ (and may be the same); and $\bar{X}_1, \bar{X}_2$ are lists of attribute names of the same length that are {\em pairwise comparable}, that is, $X_1^i$ and  $X_2^i$, and also  $\bar{Y}_1,\bar{Y}_2$, share the same domain.\footnote{A more precise notation for the MD would be: \ $\forall x_1^1 \cdots \forall y_2^m(\bigwedge_j R_1[x_1^j] \approx_j R_2[x_2^j] \ \longrightarrow \ \bigwedge_k R_1[y_1^k] \doteq R_2[y_2^k])$.}
 The MD says that, for every pair of tuples (one in relation $R_1$, the other in relation $R_2$) where the LHS is true, the attribute values in them on the RHS have to be made identical.
Symbol $\approx$ denotes generic, reflexive, symmetric, and application/domain dependent similarity relations on shared attribute domains.

 A {\em dynamic, chase-based semantics} for MDs with matching functions (MFs) was introduced in \cite{Bertossi12}. \ignore{Inspired by \cite{FanJLM09}, a  chase step requires two
 instances: a first one where the similarities hold, and second one where the matchings (equality of attribute values) are enforced. More precisely,
for a set $\Sigma$ of MDs, a pair of instances $(D,D')$ satisfies $\Sigma$, denoted $(D,D') \models \Sigma$,
if whenever $D$ satisfies the antecedents of all the MDs in $\Sigma$, then  $D'$ satisfies their consequents, with $\doteq$ interpreted  as
equality ($=$).
If $(D,D) \not \models \Sigma$, we say that $D$ is ``unresolved" (wrt. $\Sigma$). Otherwise,
$D$ is {\em stable} and considered to be duplicate-free \ \cite{FanJLM09}.} Given an initial instance $D$, the set $\Sigma$ of MDs is iteratively enforced until they cannot be be applied any further, at which point
a {\em resolved instance} has been produced.
In order to {\em enforce} (the RHSs of) MDs,  there are binary {\em matching functions} (MFs)
$\match_A: {\it Dom}_{\!A}\times {\it Dom}_{\!A}\rightarrow {\it Dom}_{\!A}$; and
$\match_A(a,a')$ is used to replace two values $a, a' \in {\it Dom}_{\!A}$ that have to be made identical.
MFs are idempotent, commutative, and associative, and then induce a partial-order
structure $\langle {\it Dom}_{\!A}, \preceq_A\rangle$, with: $a \preceq_A a' :\Leftrightarrow \match_A(a,a') = a'$ \ \cite{icdt11,BenjellounGMSWW09}.
It always holds: \ $a,a' \preceq_A \match_A(a,a')$.
  In this work, MFs are treated as built-in relations.

\ignore{
A chase-based semantics for data cleaning (or entity resolution) with MDs is given in ~\cite{icdt11}: starting from an instance $D_0$, we identify pairs of records $r_1,r_2$ that satisfy the similarity conditions on the left-hand side of a matching dependency $\varphi$, i.e., $r_1^{D_0}[\bar{X}_1] \approx r_2^{D_0}[\bar{X}_2]$ (but not its RHS), and apply an MF on the values for the right-hand side attribute,  $r_1^{D_0}[A_1],r_2^{D_0}[A_2]$, to make them both equal to $\match_A(r_1^{D_0}[A_1],r_2^{D_0}[A_2])$. We keep doing this on the resulting instance, in a chase-like procedure, until a stable instance is reached. }

\ignore{
We briefly recall the chase semantics mentioned above \cite{Bertossi12}. Consider database instances $D,D'$ with the same record identifiers,
$\Sigma$ a set of MDs, and $\varphi \in \Sigma$ of the form
$\varphi\!: \ R_1[\bar{X}_1] \approx R_2[\bar{X}_2] \rightarrow R_1[\bar{Y}_1] \doteq R_2[\bar{Y}_2]$; and
$r_1,r_2$ be an $R_1$-record and an $R_2$-record identifiers, respectively.
We say that $D'$ is the {\em immediate result of enforcing} $\varphi$ on $r_1,r_2$ on
$D$, denoted $(D,D')_{[r_1,r_2]} \models \varphi$, if: \ (a) $r_1^D[\bar{X}_1] \approx r_2^D[\bar{X}_2]$, but $r_1^D[\bar{Y}_1] \neq r_2^D[\bar{Y}_2]$.
\ (b) $r_1^{D'}[\bar{Y}_1] = r_2^{D'}[\bar{Y}_2] = \match_A(r_1^D[\bar{Y}_1], r_2^D[\bar{Y}_2])$. \ (c) $D,D'$ agree on every other record and attribute value.

Now, for an instance $D_0$ and a set of MDs $\Sigma$, an instance $D_k$ is
$(D_0,\Sigma)\mbox{\em -resolved}$ if $D_k$ is stable,
and there exists a finite sequence of
instances $D_1,\ldots, D_{k-1}$ such that, for every $i \in [1,k]$,
$(D_{i-1},D_i)_{[r_1^i,r_2^i]} \models \varphi$, for some $\varphi \in \Sigma$ and record
identifiers $r_1^i,r_2^i$.
}

There  may be several resolved instances for $D$ and $\Sigma$.
However, when (a) MFs are similarity-preserving (i.e.,
$a \approx a'$ implies $a \approx \match_A(a',a'')$); or (b) $\Sigma$ is interaction-free
(i.e., each attribute may appear in either the RHS or LHS of MDs in $\Sigma$),
there is a unique resolved instance that is computable in polynomial time in  $|D|$ \cite{Bertossi12}.

\vspace{-3mm}
\subsection {Support vector machines}

\vspace{-2mm}
The SVMs technique \cite{Vapnik98} is a form of  kernel-based learning.  \ SVMs can be used for classifying vectors in  an inner-product vector space $\mc{V}$
over $\mathbb{R}$. Vectors are classified
 in two classes, with a label in $\{0,1\}$. The algorithm
learns from a training set, say  $\{(\mathbf{e}_1, f(\mathbf{e}_1)), (\mathbf{e}_2,$ $f(\mathbf{e}_2)), (\mathbf{e}_3, f(\mathbf{e}_3)),$ $\ldots, (\mathbf{e}_n,$ $f(\mathbf{e}_n))\}$. Here,
  $\mf{e}_i \in \mc{V}$, and for the {\em feature} (function) $f$: \ $f(\mf{e}_i) \in \{0, 1\}$.

SVMs find an optimal hyperplane, $\mc{H}$, in $\mc{V}$ that  separates the  two classes where the training vectors are classified. Hyperplane $\mc{H}$ has an equation of the form $\mathbf{w} \bullet  \mathbf{x} + b$, where
$\bullet$ denotes the inner product, $\mf{x}$ is a vector variable, $\mf{w}$ is a weight vector of real values, and $b$ is a real number.
Now, a new vector $\mf{e}$ in $\mc{V}$ can be classified as positive or negative depending on the side of $\mc{H}$ it lies. This is determined by computing $h(\mathbf{e}) := \nit{sign}(\mathbf{w} \bullet  \mathbf{e} + b)$. 
If $h(\mathbf{e})> 0$, $\mathbf{e}$ belongs to class $1$; otherwise, to class $0$.

\begin{wrapfigure}[7]{r}{0.40\textwidth}
\begin{center}
\vspace*{-10mm}
\begin{center}
\includegraphics[width=4.5cm]{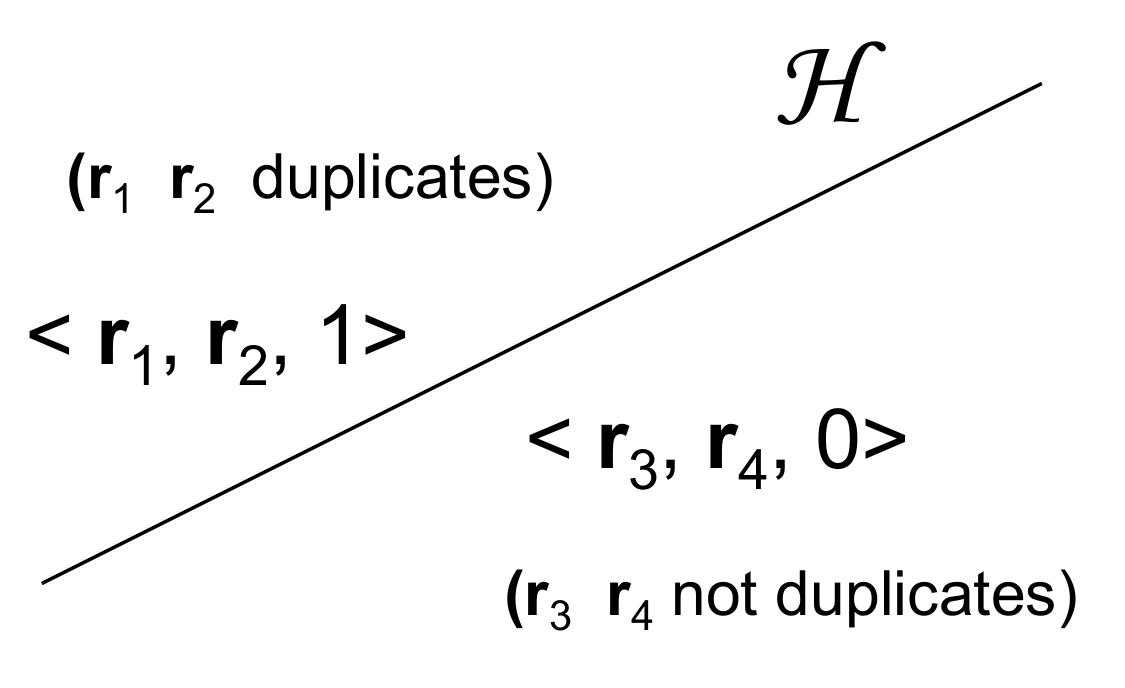}
\end{center}
\vspace{-4mm}
\caption{~~Classification hyperplane}\label{fig:sep}
\end{center}
\end{wrapfigure}

It is possible to compute real numbers $\alpha_1, \ldots, \alpha_n$, such that the classifier $h$ can be computed through:  $h(\mathbf{e}) = \nit{sign}(\sum_i \alpha_i \cdot f(\mathbf{e}_i) \cdot \mathbf{e}_i \bullet \mathbf{e} + b)$ \ (cf. Figure \ref{fig:sep}).

\ignore{
The corresponding decision function is called a classifier. In the case where the training data is linearly separable, computing an SVM for the data corresponds to minimizing $||\mathbf{w}||$, or equivalently, minimizing $\mathbf{w} \bullet \mathbf{w}$ such that

$$\mbox{For every} \ \mathbf{e} \in E, \ \ \ f(\mathbf{e}) \cdot (\mathbf{w} \bullet \mathbf{e} + b) \geq 1$$
A new vector (data point) $\mathbf{e}$ is classified by computing its value $h(\mathbf{e})$:

}

\vspace{-3cm}
\begin{figure}[h]
\hspace*{0.3cm} \includegraphics[width=12cm]{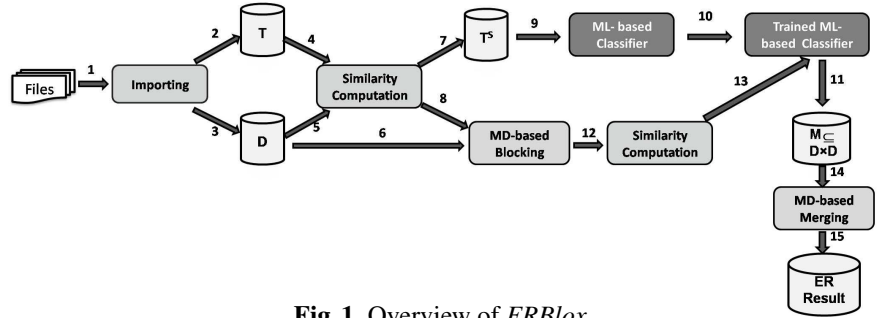}\vspace{-3cm}
  \caption{Overview of {\em ERBlox}}\label{fig:ERBlox}\vspace{-6mm}
\end{figure}

\vspace{-3mm}

\section{Overview of {\em ERBlox}}\label{MLMDFramework}

\vspace{-2mm}

A high-level description of the components of {\em ERBlox} is given in Figure \ref{fig:ERBlox}. It shows the workflow supported by {\em ERBlox} when doing ER. {\em ERBlox}'s three main components are: (1) MD-based collective blocking (path $\mf{1,3,5,\{6,8\}}$), (2) ML-based record duplicate detection (the whole initial workflow up to task $\mf{13}$, inclusive), and (3) MD-based merging (path $\mf{14, 15}$). In the figure, all the boxes in light grey are supported by {\em LogiQL}. As just done, in the rest of this section, numbers in boldface refer to the edges in this figure.


The initial input data is stored in structured text files.\\ (We assume these data are already standardized and
free of\\ misspellings, etc., but duplicates may be present.)   Our general {\em LogiQL} program that supports the whole workflow contains some rules for importing data from the files into the extensions of relational predicates (think of tables, this is  edge $\mf{1}$).  This results in a relational database instance $T$ containing the training data (edge $\mf{2}$), and the instance $D$ on which ER will be performed (edge $\mf{3}$).


   \begin{wrapfigure}[9]{r}{0.45\textwidth}
\begin{center}
\vspace*{-6mm}
\begin{center}
\includegraphics[width=5.5cm]{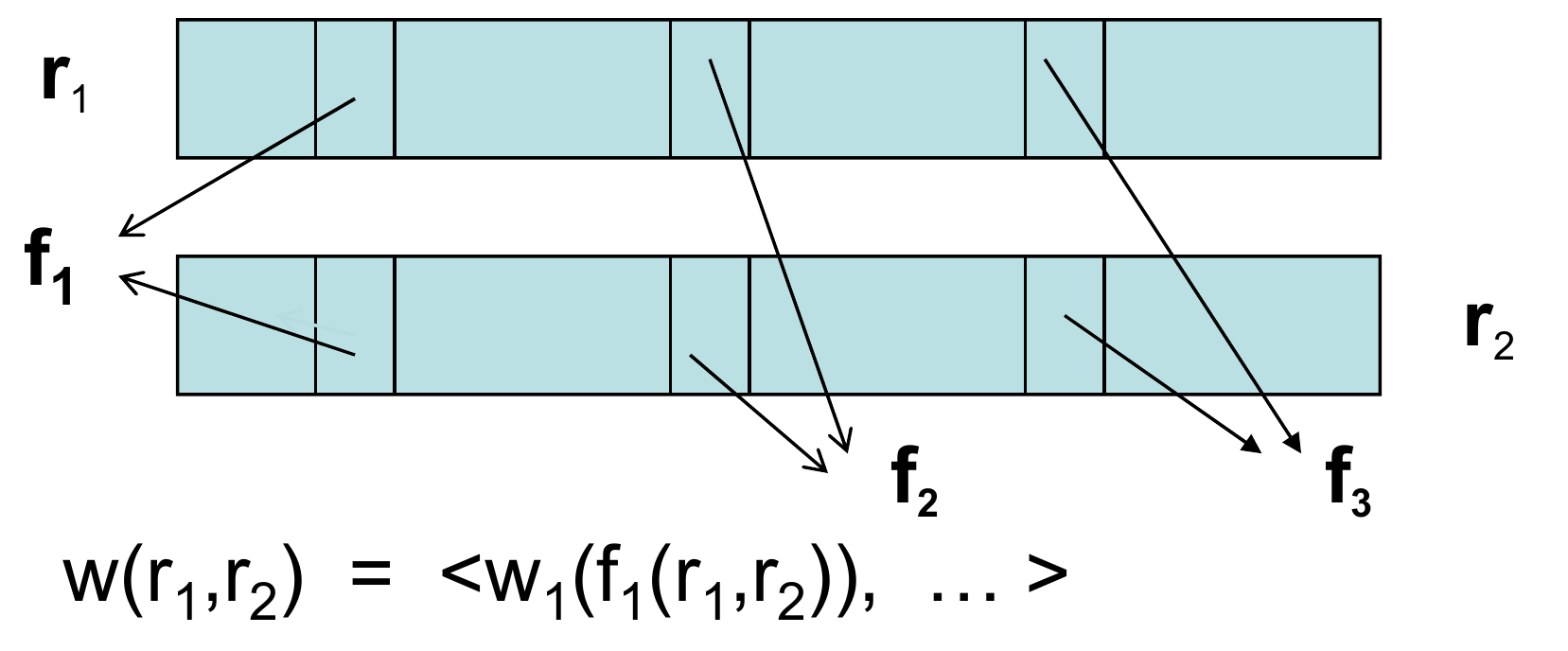}
\end{center}
\vspace{-4mm}
\caption{Feature-based similarity}\label{fig:sim}
\end{center}
\end{wrapfigure}

\ignore{\vspace{-4mm}}The next main task is blocking, which requires similarity computation of pairs of records in $D$ (edge $\mf{5}$). For record pairs $\langle r_1, r_2\rangle$ in $T$, similarities have to be computed as well (edge $\mf{4}$).  Similarity computation is based on similarity functions,
$\nit{Sf}_i\!: \nit{Dom}_{A_i} \times \nit{Dom}_{A_i} \rightarrow [0,1]$, each of which assigns a numerical value, called similarity weight, to the comparisons of values for a record attribute $A_i$ (from a pre-chosen subset of attributes) \ (cf. Figure \ref{fig:sim}). A weight vector \ $w(r_1,r_2) = \langle \cdots, \nit{Sf}_i(r_1[A_i],r_2[A_i]), \cdots \rangle$
is formed by similarity weights (edge $\mf{7}$). For more details on similarity computation see Section \ref{sec:init}.

Since  some pairs in $T$ are considered to be duplicates and others non-duplicates, the result of this process leads to a ``similarity-enhanced" database $T^{\!s}$ of tuples of
the form $\langle r_1, r_2, w(r_1,r_2), L\rangle$, with label $L \in \{0,1\}$ indicating if the two records are duplicates ($L = 1$) or not ($L=0$). The labels are consistent with the corresponding weight vectors. The classifier is trained using $T^{\!s}$, leading to a classification model (edges $\mf{9,10}$).

For records in $D$, similarity measures are needed for blocking, to decide if two records $r_1,r_2$ go to the same block. Initially, every record has its rid assigned as
block (number). To assign two records to the same block, we use matching dependencies that specify and enforce (through their RHSs) that their blocks have to be identical.
This happens when certain similarities between pairs of attribute values appearing in the LHSs of the MDs hold.
For this reason, similarity computation is also needed before blocking  (workflow $\mf{5,6,8}$). This similarity computation process
is similar to the one for $T$. However, in the case of $D$, this does not lead directly to the same kind of weight vector computation. Instead, the computation of similarity measures is only for the
 similarity predicates appearing in the LHSs of the blocking-MDs. (So, as the evaluation of the LHS in (\ref{eq:md}) requires the computation
 of similarities for $\nit{dept}$-string values.)

Notice that these blocking-MDs may capture semantic knowledge, so they could involve in their LHSs similarities of attribute values in records for different kinds of
 entities. For example, in relation to Example \ref{ex:IntroBlock}, there could be similarity comparisons involving  attributes for entities \ent{Author} and \ent{Paper}, e.g.\vspace{-2mm}
{\small\begin{eqnarray}
\nit{Author}(x_1,y_1,\nit{bl}_1) &\wedge& \nit{Paper}(y_1,z_1, \nit{bl}_3) \wedge \nit{Author}(x_2,y_2,\nit{bl}_2) \wedge \nonumber\\
 &&\hspace*{0.5cm}\nit{Paper}(y_2,z_2, \nit{bl}_4) \wedge x_1 \approx_1 x_2 \wedge z_1 \approx_2 z_2 \  \rightarrow \  \nit{bl}_1 \doteq \nit{bl}_2,\label{eq:blmd}
 \end{eqnarray}}

 \vspace{-6mm} \noindent
 expressing that when the similarities on the LHS hold, the blocks $\nit{bl}_1, \nit{bl}_2$ have to be made identical.\footnote{These MDs are more general than those
 introduced in Section \ref{sec:mds}: they may contain regular database atoms, which are used to give context to the similarity atoms in the same antecedent.}
  The similarity comparison atoms on the LHS are considered to be true when the similarity values are above  predefined thresholds (edges $\mf{5,8}$).\footnote{At this point, since all we want is to do blocking, and not yet decisions about duplicates, we could, in comparison with what is done with pairs in $T$, compute less similarity measures and and even with  low thresholds.}

This is the {\em MD-based collective blocking} stage that results in database $D$ enhanced with information about the blocks to which the records are assigned. Pairs of records with the same block form  {\em candidate duplicate record pairs}, and any two records with different blocks are simply not tested as possible duplicates (of each other).

\ignore{\red{Notice that we have similarity computation before blocking (edges $\mf{5,8}$) and after blocking (edge $\mf{12}$). These two have the same similarity functions, i.e. do the same work, except that before blocking we do not make weight vectors.  What we need is to know if attributes values are similar for enforcing the MDs. However, after blocking, we need to form weight vectors for candidate duplicate record-pairs, similar to making weight vectors for $T$. }  }

\ignore{\vspace{-3mm}}After the records have been assigned to blocks, pairs of records $\langle r_1, r_2\rangle$ in the same block are considered for the duplicate test. As this point we proceed as we did
for $T$: the similarity vectors $w(r_1,r_2)$ have to be computed (edges $\mf{11, 12}$).\footnote{Similarity computations are kept in appropriate program predicates. So similarity values computed before blocking can be reused at this stage, or whenever needed.}  Next, tuples $\langle r_1, r_2,w(r_1,r_2)\rangle$ are used as input for the trained classification algorithm (edge $\mf{12}$).

\vspace{5mm}
The result of the trained ML-based classifier, in this case obtained through SVMs as a separation hyperplane $\mc{H}$, is a set  $M$  of record pairs $\langle r_1,r_2,1\rangle$ that come from the same block and are
considered to be duplicates (edge $\mf{13}$).\footnote{The classifier also returns pairs or records that come from the same block, but are not considered to be duplicate. The set thereof in not
 interesting, at least as a workflow component.} The records in these pairs will be merged on the basis of an {\em ad hoc} set of MDs (edge $\mf{15}$), different from those used
 in edges $\mf{6,8}$.

Informally, the merge-MDs are of the form: $r_1 \approx r_2 \ \rightarrow \ r_1 \doteq r_2$, where the antecedent is true when $\langle r_1, r_2,1\rangle$ is an output of
the classifier. The RHS is a shorthand for: \ $r_1[A_1] \doteq r_2[A_1] \wedge \cdots \wedge r_1[A_m] \doteq r_2[A_m]$, where $m$ is the total number of record attributes. Merge at
the attribute level uses the  matching functions $\match_{A_i}$.

We point out that MD-based merging takes care of transitive cases provided by the classifier, e.g. if it returns  $\langle r_1, r_2,1\rangle$, $\langle r_2, r_3,1\rangle$, but not $\langle r_1, r_3,1\rangle$, we still merge $r_1, r_3$ (even when $r_1 \approx r_3$  does not hold). Actually, we do this by by merging all the records $r_1,r_2, r_3$ into the same record. Our system is capable of recognizing this situation and solving it as expected. This relies on the way we store and manage -via our {\em LogiQL} program- the positive cases obtained
from the classifier (details can be found in Section \ref{Merging}). In essence, this makes our set of merging-MDs {\em interaction-free}, and leads to a  unique resolved instance
\cite{Bertossi12}.

\ignore{
\comlb{READ CAREFULLY: I think something should be said in general terms about the ``transitive case", i.e. a situation like $\langle r_1, \red{r_2},1\rangle$, $\langle \red{r_2}, r_3,1\rangle$. (I do not see why
the classifier should output $\langle r_1, r_3,1\langle$, its does not have to be ``duplicate-preserving" or transitive). How is the overall merging for all
the records that are transitively connected considering that MDs are applied to pairs or records? After merging $r_1,r_2$, will the connection $r_2,r_3$ be lost? Could there be more than one
resolved instance depending on the order of application of MDs? The ``non-interaction" by itself will take care of this. Etc. Even assuming similarity-preserving MFs does not work, because
the condition for application of an MD is that the record pair is a positive output for the classifier. So, the the tuple $\langle r_1,r_3\rangle$ may not be an output of the classifier even when
$r_1 \approx' r_3$ at the attribute level. This says that, *before applying the MDs*, it would be necessary ``to produce the transitive closure" the output
of the classifier. This is all rather blurred and need some explaining.}
}

The following sections provide more details on {\em ERBlox}  and our approach to ER.

\vspace{-4mm}
\section{Initial Data and Similarity Computation}\label{sec:init}

\vspace{-3mm}
We describe now some aspects of the MAS dataset,  highlighting the input for- and output of each component of the {\em ERBlox} system.
The data is represented and provided as follows. The \ent{Author} relation contains authors names and their affiliations. The \ent{Paper} relation contains paper titles, years, conference IDs, journal IDs, and keywords. The \ent{PaperAuthor} relation contains papers IDs, authors IDs, authors names, and their affiliations. The \ent{Journal} and \ent{Conference} relations contain short names, full names, and home pages of journals and conferences, respectively. By using {\em ERBlox} on this dataset, we determine which papers in MAS data are written by a given author. This is clear case of ER since there are many authors who publish under several variations of their names. Also the same paper may appear under slightly different titles, etc.\footnote{For our  experiments,
 we
independently used  two other datasets: DBLP and Cora Citation.}

From the MAS dataset, which contains the data in structured files, extensions for intentional, relational predicates are computed by {\em LogiQL}-rules of the general program, e.g. \vspace{-2mm}
{\small \begin{eqnarray}
&&\_\nit{file}\_\nit{in}(x1,x2,x3) \rightarrow \nit{string}(x1),\nit{string}(x2), \nit{string}(x3). \label{eq:one}\\
&&\nit{lang:physical:filePath}[`\_\nit{file}\_\nit{in}] = "\nit{author.csv}". \label{eq:two}\\
&&\nit{+author}(\nit{id}1,x2,x3) \leftarrow \_\nit{file}\_\nit{in}(x1, x2,x3),\nit{string\!:\!int}64\!:\!\nit{convert}[x1] = \nit{id}1. \label{eq:three}
\end{eqnarray}}

\vspace{-6mm} \noindent
Here, (\ref{eq:one}) is a predicate schema declaration (metadata uses ``$\rightarrow$"), in this case of  the ``$\_$\ent{file}$\_$\ent{in}" predicate with three string-valued attributes,\footnote{In {\em LogiQL}, each predicate has to be declared,  unless it can be inferred from the rest of the program.} which is used to store the contents extracted from the source file, whose path is specified by (\ref{eq:two}).
Derivation rules, such as (\ref{eq:three}), use the usual ``$\leftarrow$". In this case, it defines the \ent{author} predicate, and the ``$+$" in the rule head inserts the data into the predicate extension.
The first attribute is made an identifier \cite{Aref15}. Figure \ref{fig:dataset} illustrates a  small part of the dataset obtained by importing data into the relational  predicates. (There may be missing attributes values.)

\begin{figure}[t]
{\scriptsize
\begin{center}\begin{tabular}{r|r|l|l|r|}\hline
$\nit{Author}$& $\nit{AID}$ &$\nit{Name}$ & $\nit{Affiliation}$ & $Bl \#$  \\ \hline
$$ &$659$&$\nit{Jean}\mymathhyphen{Pierre\ Olivier\ de}$ & $\nit{Ecole\ des\ Hautes}$ & $659$  \\
$$ &$2546$&$\nit{Olivier\ de\ Sardan}$ & $\nit{Recherche\ Scientifique}$ & $2546$  \\
$$ &$612$&$\nit{Matthias\ Roeckl}$ & $\nit{German\ Aerospace\ Center}$ & $612$ \\
$$ &$4994$&$\nit{Matthias\ Roeckl}$ & $\nit{Institute\ of\ Communications}$ & $4994$ \\\cline{2-5}
\end{tabular}\end{center}
\begin{center}\begin{tabular}{r|r|l|c|c|c|l|c|}\hline
$\nit{Paper}$&  $\nit{PID}$ &$\nit{Title}$ & $\nit{Year}$  & $CID$ & $JID$&$Keyword$& $Bl \#$\\ \hline
$$ &$123$&$\nit{Illness\ entities\ in\ West\ Africa}$ & $1998$ & $179$& $$ & $\nit{West\ Africa,\ Illness}$ & $123$\\
$$ &$205$&$\nit{Illness\ entities\ in\ Africa}$ & $1998$  & $179$ &  $$ &$\nit{Africa, Illness}$& $205$\\
$$ &$769$&$\nit{DLR\ Simulation\ Environment\ m3}$ & $2007$  & $146$ &  $$ &$\nit{ Simulation\ m3}$& $769$\\
$$ &$195$&$\nit{DLR\ Simulation\ Environment}$ & $2007$  & $146$ &  $$ &$\nit{ Simulation}$& $195$\\\cline{2-8}
\end{tabular}\end{center}
\begin{center}\begin{tabular}{c|c|r|l|l|}\hline
$\nit{PaperAuthor}$&$\nit{PID}$ & $\nit{AID}$ & $\nit{Name}$ & $\nit{Affiliation}$ \\ \hline
$$&$123$ & $659$ & $\nit{Jean}\mymathhyphen\nit{Pierre\ Olivier\ de}$ & $\nit{Ecole\ des\ Hautes}$  \\
$$&$205$ & $2546$ & $\nit{Olivier\ de\ Sardan}$ & $\nit{Recherche\ Scientifique}$ \\
$$&$769$ & $612$ & $\nit{Matthias\ Roeckl}$ & $\nit{German\ Aerospace\ Center}$ \\
$$&$195$ & $4994$ & $\nit{Matthias\ Roeckl}$ & $\nit{Institute\ of\ Communications}$ \\ \cline{2-5}
\end{tabular}\end{center}
\ignore{\begin{center}\begin{tabular}{r|r|l|l|l|r|}\hline
$\nit{Journal}$& $\nit{CID}$ &$\nit{SName}$ & $\nit{FName}$ & $\nit{HPage}$ & $\nit{Bl\#}$ \\ \hline
$$ &$189$&$\nit{F.\ Cass}$ & $\nit{Frank\ Cass}$ & $$ & $189$  \\
$$ &$152$&$\nit{INTERNET}$ & $\nit{IEEE\ Internet\ Computing}$ & $\nit{computer.org/internet}$ & $152$ \\ \cline{2-6}
\end{tabular}\end{center}
\begin{center}\begin{tabular}{r|r|c|l|c|c|}\hline
$\nit{Conference}$& $\nit{CID}$ &$\nit{SName}$ & $\nit{FName}$ & $\nit{HPage}$ & $\nit{Bl\#}$  \\ \hline
 $$ &$179$&$$ & $\nit{Medical\ Anthropology }$ & $\nit{medant.com}$ & $179$ \\
 $$ &$146$&$$ & $\nit{First\ C2C}\mymathhyphen\nit{CC}\mymathhyphen\nit{COMeSafety\ Simulation}$ & $$ & $146$ \\ \cline{2-6}
\end{tabular}\end{center}
\begin{center}\begin{tabular}{c|l|l|}\hline
$\nit{CoAuthor}$&$\nit{AID}_1$ & $\nit{AID}_2$  \\ \hline
$$&$5026$ & $659$  \\
$$&$5026$ & $2546$ \\ \cline{2-3}
\end{tabular}\end{center}}}
\vspace{-4mm}\caption{Relation extensions from MAS using  LogiQL rules} \label{fig:dataset} \vspace{-7mm}
\end{figure}

 As described above, in {\em ERBlox}, similarity computation generates similarity weights, which are used to: (a) compute the weight vectors for the training data $T$ and the data in $D$
 under classification; and (b) do the blocking, where  similarity weights are compared with predefined thresholds for the similarity conditions  in the LHSs of blocking-MDs.\footnote{As described at the end of Section \ref{MLMDFramework}, these similarity computations are {\em not} used with the MDs that support the final merging process (cf. Section \ref{Merging}).}

We used three well-known similarity functions \cite{cohen03}, depending on the  attribute domains. ``TF-IDF cosine similarity" \cite{Salton88} used for computing similarities for text-valued attributes, whose values are string vectors. It assigns low weights to frequent strings and high weights to rare strings. It was used for attribute values that contain frequent strings, such as affiliation.
For attributes with short string values, such as author name, we applied ``Jaro-Winkler similarity" \cite{Winkler99}. Finally, for numerical attributes, such as publication year, we used ``Levenshtein distance" \cite{Navarro}, which computes similarity of two numbers on the basis of the minimum number of operations required to transform one into the other.

Similarity computation for {\em ERBlox}
is supported by {\em LogiQL}-rules that define similarity functions. In particular, similarity computations are kept in extensions of program predicates. For example, if the similarity weight of  values $a_1,a_2$ for
attribute $\nit{Title}$ is above the threshold, a tuple $\nit{TitleSim}(a_1,a_2)$ is created by  the program.

\vspace{-3mm}
\section{MD-Based Collective Blocking and Duplicate Detection} \label{MDBlocking}

\vspace{-2mm}
Since every record has an identifier, {\it rid}, initially each record uses its \nit{rid} as its block number, in an extra attribute  $\nit{Bl\#}$.  In this way, we create the {\em initial blocking instance} from the initial instance $D$, also denoted with $D$. Now, blocking strategies are captured by means of (blocking) MDs of the form: \vspace{-2mm}
\begin{equation}
R_i(\bar{X}_1,\nit{Bl}_1) \wedge R_i(\bar{X}_2,\nit{Bl}_2) \wedge \psi(\bar{X}_3) \ \to \ \nit{Bl}_1\doteq \nit{Bl}_2. \label{eq:newMDs}
\end{equation}

\vspace{-3mm}\noindent
Here $\nit{Bl}_1, \nit{Bl}_2$ are variables for block numbers, and $R_i$ is a database (record) predicate. The lists of variables $\bar{X}_1, \bar{X}_2$ stand for all the attributes in $R_i$, but $\nit{Bl\#}$. Formula $\psi$ is a conjunction of relational atoms and comparison atoms via similarity predicates;  but it does not
contain similarity comparisons of blocking numbers, such as  $\nit{Bl}_3\!\approx \nit{Bl}_4$.\footnote{Actually, this natural condition makes the set of blocking-MDs interaction-free, i.e. for every two blocking-MDs $m_1, m_2$, the set of attributes on the RHS of $m_1$ and the set of attributes on the LHS of $m_2$ on which there are similarity predicates, are disjoint \cite{Bertossi12}.}  The variables in the list $\bar{X}_3$ appear in $R_i$ or in another database predicate or in a similarity atom. It holds that $(\bar{X}_1 \cup \bar{X}_2) \cap \bar{X}_3 \neq\emptyset$. For an example, see (\ref{eq:blmd}), where $R_i$ is \ent{Author}.

In order to enforce these MDs on two records, we use a binary matching function $\match_{\!_\nit{Bl\#}}$, to make two block numbers identical: $\match_{\!_\nit{Bl\#}}(i, j):= i \ \ \mbox{if} \ \ j \leq i$.
More generally, for the application-dependent set, $\Sigma^{\nit{Bl}}$, of blocking-MDs we adopt the  chase-based semantics for entity resolution \cite{Bertossi12}. Since this set of MDs is interaction-free, its enforcement results in a single  instance $D^{\!\nit{Bl}}$,  where now records may share block numbers, in which case  they belong to the the same block. Every record is assigned to a single block.

\vspace{-1mm}
\begin{example}\label{ex:blockLoqic} These are some of the blocking-MDs used for the MAS dataset:

\vspace{-0.45cm}
{\scriptsize \begin{eqnarray}
&&\hspace*{-0.8cm}  \nit{Paper}(\nit{pid}_1,x_1,y_1,z_1,w_1,v_1,\nit{bl}_1) \wedge \nit{Paper}(\nit{pid}_2,x_2,y_2,z_2,w_2,v_2,\nit{bl}_2) \wedge \label{eqq:md1}\\
&& \hspace*{4.0cm} \ x_1 \approx_{\!\nit{Title}} x_2 \ \wedge \ y_1=y_2 \ \wedge z_1=z_2 \to \ \nit{bl}_1 \doteq \nit{bl}_2. \nonumber\\
&&\hspace*{-0.8cm}  \nit{Author}(\nit{aid}_1, x_1,y_1, \nit{bl}_1) \ \wedge \ \nit{Author}(\nit{aid}_2, x_2,y_2, \nit{bl}_2) \wedge \label{eqq:md2} \\
&&\hspace*{4.9cm} \ x_1 \approx_{\!\nit{Name}} x_2 \ \wedge \ y_1\approx_{\!\nit{Aff}}y_2 \to \ \nit{bl}_1 \doteq \nit{bl}_2.\nonumber  \\
&&\hspace*{-0.8cm}  \nit{Paper}(\nit{pid}_1,x_1,y_1,z_1,w_1,v_1,\nit{bl}_1) \ \wedge \ \nit{Paper}(\nit{pid}_2,x_2,y_2,z_2,w_2,v_2,\nit{bl}_2) \wedge \label{eqq:md3} \\
&&\hspace*{0.30cm}\nit{PaperAuthor}(\nit{pid}_1, \nit{aid}_1, x'_1,y'_1) \ \wedge \nit{PaperAuthor}(\nit{pid}_2, \nit{aid}_2, x'_2,y'_2) \wedge \nonumber \\
&&\hspace*{1cm} \nit{Author}(\nit{aid}_1, x'_1,y'_1, \nit{bl}_3) \ \wedge \nit{Author}(\nit{aid}_2, x'_2,y'_2, \nit{bl}_3) \wedge x_1 \approx_{\!\nit{Title}} x_2   \to \ \nit{bl}_1 \doteq \nit{bl}_2.\nonumber \\
&&\hspace*{-0.8cm} \nit{Author}(\nit{aid}_1, x_1,y_1, \nit{bl}_1) \ \wedge \nit{Author}(\nit{aid}_2, x_2,y_2, \nit{bl}_2) \ \wedge \ x_1 \approx_{\!\nit{Name}} x_2\wedge  \label{eqq:md4}\\
&&\hspace*{0.35cm}
 \nit{PaperAuthor}(\nit{pid}_1, \nit{aid}_1, x_1,y_1) \ \wedge \nit{PaperAuthor}(\nit{pid}_2, \nit{aid}_2, x_2,y_2) \wedge \nonumber\\
 &&\hspace*{4mm}\nit{Paper}(\nit{pid}_1,x'_1,y'_1,z'_1,w'_1,v'_1,\nit{bl}_3)  \wedge  \nit{Paper}(\nit{pid}_2,x'_2,y'_2,z'_2,w'_2,v'_2,\nit{bl}_3) \to \ \nit{bl}_1 \doteq \nit{bl}_2.\nonumber
\ignore{&m_5\!:&  \nit{CoAuthor}(\nit{aid}_1,\nit{aid}_3) \ \land \nit{CoAuthor}(\nit{aid}_2,\nit{aid}_3) \ \wedge \  x_1\approx_{\!\nit{Name}}  x_2 \wedge \nonumber  \\ &&\hspace*{0.40cm} \nit{Author}(\nit{aid}_1, x_1,y_1,\nit{bl}_1) \ \land \ \nit{Author}(\nit{aid}_2, x_2,y_2,\nit{bl}_2)\ \to \nit{bl}_1 \doteq \nit{bl}_2.\label{eqq:md5}}
\end{eqnarray}}

\vspace{-6mm}
\noindent Informally, (\ref{eqq:md1}) tells us that, for every two \ent{Paper} entities $\mathbf{p}_1, \mathbf{p}_2$  for which the values
for attribute $\nit{Title}$ are similar and with same publication year, conference ID,  the values for attribute
${\it Bl\#}$ must be made the same. By (\ref{eqq:md2}), whenever there are similar values for name and affiliation in \ent{Author}, the corresponding authors should be in the same block. Furthermore, (\ref{eqq:md3}) and (\ref{eqq:md4}) collectively block \ent{Paper} and \ent{Author} entities. For instance,  (\ref{eqq:md3}) states that if two authors are in the same block, their papers $\mathbf{p}_1$, $\mathbf{p}_2$ having similar titles must be in the same block. Notice that if papers $\mathbf{p}_1$ and $\mathbf{p}_2$ have similar titles, but they do not have same publication year or conference ID, we cannot block them together using (\ref{eqq:md1}) alone. \boxtheorem
\end{example}
\vspace{-2mm}We now show how these MDs are represented in {\em LogiQL}, and how we use {\em LogiQL} programs for declarative specification of MD-based collective blocking.\footnote{Notice that since we have interaction-free sets of blocking-MDs, stratified Datalog programs are expressive enough to express and enforce them \cite{Bahmani12}. {\em LogiQL} supports stratified Datalog.} In {\em LogiQL}, an MD takes the form: \vspace{-2mm}
{\small \begin{equation}
R_i[\bar{X}_1]\!=\!\nit{Bl}_2 , \ \ R_i[\bar{X}_2]\!=\!\nit{Bl}_2  \ \ \longleftarrow \ \ \ R_i[\bar{X}_1]=\nit{Bl}_1, \ R_i[\bar{X}_2]=\nit{Bl}_2, \  \psi(\bar X_3), \ \nit{Bl}_1 < \nit{Bl}_2,\label{LQmd}
\end{equation} }

\vspace{-5mm}\noindent subject to the same conditions as in (\ref{eq:newMDs}).  An atom $R_i[\bar{X}]\!\!=\!\!\nit{Bl}$ states  that predicate $R_i$ is functional on $\bar{X}$ \cite{Aref15}.  It means each record in $R_i$ can have only one block number  $\nit{Bl\#}$. 

\ignore{\begin{example} \label{blockMDtoLQ} (example \ref{ex:blockLoqic} cont.) For instance, the MD (\ref{eqq:md5}) is represented in {\em LogiQL}  by the following rules:
{\small \begin{eqnarray*}
&&  \nit{Author}[\nit{aid}_1, x_1,y_1]=\nit{bl}_2 \ , \nit{Author}[\nit{aid}_2, x_2,y_2]=\nit{bl}_2 \leftarrow \nit{CoAuthor}(\nit{aid}_1,\nit{aid}_3),  \\ &&\hspace*{1.0cm} \nit{CoAuthor}(\nit{aid}_2,\nit{aid}_3), \nit{Author}[\nit{aid}_1, x_1,y_1]=\nit{bl}_1 , \nit{Author}[\nit{aid}_2, x_2,y_2]=\nit{bl}_2, \\ &&\hspace*{7.50cm}\  \nit{NameSim}(x_1, x_2) \ , \nit{bl}_1 < \nit{bl}_2.
\end{eqnarray*}}\boxtheorem
\end{example}}

Given an initial instance $D$, a {\em LogiQL} program $\mc{P}^{B}(D)$ that specifies MD-based collective blocking contains the following (kind of) rules:

\vspace{2mm}
\noindent {\bf 1.}~
For every atom $R(\nit{rid}, \bar{x}, \nit{bl})$ $\in$ $D$, the fact  $R[\nit{rid}, \bar{x}]=\nit{bl}$. \ (Initially, $\nit{bl} := \nit{rid}$.)

\vspace{2mm}
\noindent {\bf 2.}~
For every attribute $A$ of $R_i$, facts of the form $\mbox{\nit{A-Sim}}(a_1,a_2)$, with
$a_1, a_2 \in \nit{Dom}_{\!A}$, the finite attribute domain. They are obtained by similarity computation.

\vspace{2mm}
\noindent {\bf 3.}~ The blocking-MDs as in  (\ref{LQmd}).

\ignore{$m_j\!\!: R_i[r_1,\bar{x}_1]\!\!=\!\nit{bl}_2, \ R_i[r_2,\bar{x}_2]\!\!=\!\nit{bl}_2  \leftarrow  \psi(\bar x_3), R_i[r_1,\bar{x}_1]\!=\!\nit{bl}_1 , R_i[r_2,\bar{x}_2]\!=\!\nit{bl}_2, \nit{bl}_1 < \nit{bl}_2$,  the rule:   \vspace{-5mm}
{\small\begin{eqnarray*}&&R_i[r_1,\bar x_1]\!=\! \nit{bl}_2, \ R_i[r_2,\bar x_2]\!=\! \nit{bl}_2 \;\; \leftarrow
\; R_i[r_1,\bar x_1]\!=\!\nit{bl}_1,\; R_i[r_2,\bar x_2]\!=\!\nit{bl}_2, \psi(\bar x_3),\; \nit{bl}_1 < \nit{bl}_2.
\end{eqnarray*}}
\noindent This rule is used to enforce the MD $m_j$ whenever its LHS hold. \red{The larger blocking number is used for making two blocking numbers identical.}
}
\ignore{Here, $\psi$ is conjunctions of atoms and similarity predicates. $\bar{x}_1$ is a list of variables of relation $R$ except $\nit{Bl\#}$; similarly for $\bar{x}_2. \bar{x}_3$ is a list of variables of relations in $\psi$ where $\bar{x}_1, \bar{x}_2 \cap \bar{x}_3 \neq\emptyset$.}

\vspace{2mm}
\noindent {\bf 4.}~  Rules to represent the consecutive versions of entities during MD-enforcement:

\vspace{-6mm}
{\small
\begin{eqnarray*} && R\mbox{-}\nit{OldVersion}(r_1,\bar{x}_1, \nit{bl}_1)  \ \ \leftarrow \ \  R[r_1, \bar{x}_1]=\nit{bl}_1, \ R[r_1, \bar{x}_1]=\nit{bl}_2, \ \nit{bl}_1 < \nit{bl}_2.
\end{eqnarray*}
}

\vspace{-7mm}
\noindent For each {\it rid}, $r$, there could be several
atoms of the form  $R[r,\bar x]\!=\! \nit{bl}$,  corresponding to the evolution of the record identified by $r$ due to MD-enforcement. The rule specifies that
 versions of records with lower block numbers are old.

\vspace{2mm}
\noindent {\bf 5.}~ Rules that collect the latest versions of records. They are used to form blocks:

\vspace{-6mm}
{\small \begin{eqnarray*}
&&R\mbox{-}\nit{MDBlock}[r_1,\bar{x}_1]= \nit{bl}_1 \;\;\leftarrow \;\;R[r_1,\bar{x}_1]= \nit{bl}_1, \ ! \ R\mbox{-}\nit{OldVersion}(r_1,\bar{x}_1, \nit{bl}_1).
\end{eqnarray*} }

\vspace{-6mm}
\noindent  In {\em LogiQL}, \ ``!", as in the body above, is used for negation \cite{Aref15}. The rule collects $R$-records that are not old versions.

\vspace{2mm}Programs $\mc{P}^{B}(D)$ as above are stratified (there is no recursion involving negation). Then, as expected in relation to the blocking-MDs, they have a single model, which can
be used to read the final block number for each record.

\begin{example}\label{LP} (ex. \ref{ex:blockLoqic} cont.) Considering only MDs (\ref{eqq:md1}) and (\ref{eqq:md3}), the portion of $\mc{P}^{B}(D)$ for blocking \ent{Paper} entities has the following rules:

\vspace{-2mm}
{\small
\begin{itemize}

\item [{\bf 2.}] Facts such as: \
{\scriptsize$\nit{TitleSim}(\nit{Illness\ entities\ in\ West\ Africa},\nit{Illness\ entities\ in\ Africa})$.\\ \hspace*{1.8cm}$\nit{TitleSim}(\nit{DLR\ Simulation\ Environment\ m3}, \nit{DLR\ Simulation\ Environment}).$}

\vspace{2mm}
\item [{\bf 3.}] {\scriptsize $\nit{Paper}[\nit{pid}_1,x_1,y_1,z_1,w_1,v_1]=\nit{bl}_2, \nit{Paper}[\nit{pid}_2,x_2,y_2,z_2,w_2,v_2]=\nit{bl}_2 \ \leftarrow$

\hfill $\nit{Paper}[\nit{pid}_1,x_1,y_1,z_1,w_1,v_1]=\nit{bl}_1, \nit{Paper}[\nit{pid}_2,x_2,y_2,z_2,w_2,v_2]=\nit{bl}_2,$

\hfill$\nit{TitleSim}(x_1,x_2), y_1=y_2, z_1=z_2, \nit{bl}_1 < \nit{bl}_2.$}

\vspace{1mm} {\scriptsize
$\nit{Paper}[\nit{pid}_1,x_1,y_1,z_1,w_1,v_1]=\nit{bl}_2, \nit{Paper}[\nit{pid}_2,x_2,y_2,z_2,w_2,v_2]=\nit{bl}_2\ \leftarrow$

 \hfill $\nit{Paper}[\nit{pid}_1,x_1,y_1,z_1,w_1,v_1]=\nit{bl}_1, \nit{Paper}[\nit{pid}_2,x_2,y_2,z_2,w_2,v_2]=\nit{bl}_2, \nit{TitleSim}(x_1,x_2),$

\hfill $ \nit{PaperAuthor}(\nit{pid}_1,\nit{aid}_1, x'_1,y'_1), \nit{PaperAuthor}(\nit{pid}_2,\nit{aid}_2, x'_2,y'_2),$

\hfill $ \nit{Author}[\nit{aid}_1, x'_1,y'_1]=\nit{bl}_3, \nit{Author}[\nit{aid}_2, x'_2,y'_2]=\nit{bl}_3, \nit{bl}_1 < \nit{bl}_2$.}

\ignore{\vspace{-7mm}
\begin{eqnarray*}
&&\hspace{0.2cm} \nit{Author}'[\nit{aid}_1, x_1,y_1]=\nit{bl}_2, \nit{Author}'[\nit{aid}_2, x_2,y_2]=\nit{bl}_2 \leftarrow \nit{Author}[\nit{aid}_1, x_1,y_1]=\nit{bl}_1 , \\&&
\hspace*{1.05cm}  \nit{Author}'[\nit{aid}_2,x_2,y_2]=\nit{bl}_2, \nit{PaperAuthor}'(\nit{pid}_1,\nit{aid}_1, x_1,y_1), \nit{NameSim}(x_1, x_2), \\&&\hspace*{1.05cm} \nit{PaperAuthor}'(\nit{pid}_2,\nit{aid}_2, x_2,y_2),
\nit{Paper}'[\nit{pid}_1,x_1,y_1,z_1,w_1,v_1]=\nit{bl}_3, \nit{bl}_1 < \nit{bl}_2,\\&&
\hspace*{7.05cm}  \nit{Paper}'[\nit{pid}_2,x_2,y_2,z_2,w_2,v_2]=\nit{bl}_3.
\end{eqnarray*}

\vspace{-7mm}
\begin{eqnarray*}
&& \hspace*{0.2cm}  \nit{Author}'[\nit{aid}_1, x_1,y_1]=\nit{bl}_2 \ , \nit{Author}'[\nit{aid}_2, x_2,y_2]=\nit{bl}_2 \leftarrow \nit{CoAuthor}'(\nit{aid}_1,\nit{aid}_3), \\ &&\hspace*{1.10cm}  \nit{CoAuthor}'(\nit{aid}_2,\nit{aid}_3), \nit{Author}'[\nit{aid}_1, x_1,y_1]=\nit{bl}_1 , \nit{Author}'[\nit{aid}_2, x_2,y_2]=\nit{bl}_2, \\ &&\hspace*{8.0cm}  \nit{NameSim}(x_1, x_2) \ , \nit{bl}_1 < \nit{bl}_2.
\end{eqnarray*}

\vspace{-2mm}   }

\vspace{1mm}
\item [{\bf 4.}] {\scriptsize $\nit{PaperOldVersion}(\nit{pid}_1,x_1,y_1,z_1,w_1,v_1,\nit{bl}_1) \! \leftarrow \! \nit{Paper}[\nit{pid}_1,x_1,y_1,z_1,w_1,v_1] =\nit{bl}_1,$

 \hfill $\nit{Paper}[\nit{pid}_1,x_1,y_1,z_1,w_1,v_1]=\nit{bl}_2, \nit{bl}_1 < \nit{bl}_2.$}

\vspace{1mm}
\item [{\bf 5.}] {\scriptsize $\nit{PaperMDBlock}[\nit{pid},\bar{x}_1]= \nit{bl}_1 \;\;\leftarrow \;\;\nit{Paper}[\nit{pid}_1,x_1,y_1,z_1,w_1,v_1]= \nit{bl}_1,$

\hfill ${\it PaperOldVersion}(\nit{pid}_1,x_1,y_1,z_1,w_1,v_1, \nit{bl}_1).$}
\end{itemize} }

\vspace{-3mm}
\noindent Restricting the model of the program to the relevant attributes of predicate \nit{PaperMDBlock} returns: $\{\{123, 205\}, \{195 ,769\}\}$, i.e. the papers with \nit{pids} $123$ and $205$ are blocked together;
similarly for those with \nit{pids} $195$ and $769$.\boxtheorem
\end{example}
As described above, the input to the trained classifier is a set of tuples of the form $\langle r_1,r_2, w(r_1,r_2) \rangle$, with $w(r_1,r_2)$ the computed weight vector for records (with ids) $r_1, r_2$ in a  same
block.\footnote{The features considered in a weight vector computation depend on whether they have a strong discrimination power, i.e. do not contain missing values.}

\begin{example}\label{ex:similarityComputation} (ex. \ref{LP} cont.) Consider the blocks for entity \ent{Paper}. If the ``journal ID" values are null in both records, but not the ``conference ID" values,  ``journal ID" is not considered
for a feature. Similarly, when the conference ID values are null. However, the values for  ``journal ID" and ``conference ID" are replaced by  ``journal full name" and ``conference full name" values,
 found in \ent{Conference} and \ent{Journal} records, resp. In this case then, attributes \nit{Title}, \nit{Year}, \nit{ConfFullName} or \nit{JourFullName} and \nit{Keyword} are used for corresponding feature for
 weight vector computation.

 Considering the previous \ent{Paper} records, the input to the classifier consists of:  $\langle 123,$ $205, w(123,205) \rangle$, with $w(123,205) =[0.8, 1.0, 1.0, 0.7]$, and $\langle 195,769, w(195,769)\rangle$, with  $w(195, 769)=[0.93, 1.0, 1.0, 0.5]$ \ignore{where the weight vectors are generated using the mentioned similarity functions for the MAS dataset. Notice that the real input to the trained classifier, such as SVM, is}
  (actually the contents of the two square brackets only). \ignore{real inputs are merely $[0.8, 1.0, 1.0, 0.7]$ and $[0.93, 1.0, 1.0, 0.5]$. }\boxtheorem
\end{example}

\vspace{-2mm}\noindent
  Several ML techniques are accessible from {\em LogicBlox} platform through the {\em BloxMLPack} library, that provides a generic Datalog interface. Then, {\em ERBlox} can call an  ML-based record duplicate detection component
  through the general {\em LogiQL} program. In this way, the SVMs package is invoked by {\em ERBlox}. \ignore{For the dataset of MAS, we apply three classifiers namely, SVM, K-NN, NBC for implementing ML-based classification of {\em ERBlox} system.}

The output  is a set of tuples of the form $\langle r_1, r_2, 1\rangle$ or $\langle r_1, r_2, 0\rangle$, where $r_1,r_2$ are ids for records of entity (table) $R$.  In the former case, a tuple $R\mbox{-}\nit{Duplicate}(r_1,r_2)$ is created (as defined by the {\em LogicQL} program). In the previous
example, the SVMs method return $\langle[0.8, 1.0, 1.0, 0.7],1\rangle$ and $\langle[0.93, 1.0, 1.0, 0.5],1\rangle$, then $\nit{PaperDuplicate}(123, 205)$ and $\nit{PaperDuplicate}(195,769)$ are created.

\vspace{-4mm}

\section{MD-Based Merging}\label{Merging}

\vspace{-3mm}
When $\nit{EntityDuplicate}(r_1,r_2)$ is created, the corresponding full records $\bar{r}_1,\bar{r}_2$ have to be merged via record-level merge-MDs
 of the form \ $R[r_1] \approx R[r_2] \ \longrightarrow \ R[\bar{r}_1] \doteq R[\bar{r}_2]$,
where $R[r_1] \approx R[r_2]$ is true when $R\mbox{-}\nit{Duplicate}(r_1,r_2)$ has been created according to the output of the SVMs classifier. The
RHS means that the two records are merged into a new full record $\bar{r}$, with $\bar{r}[A_i] := \match_{_{A_i}\!\!}(\bar{r}_1[A_i],\bar{r}_2[A_i])$ \ \cite{Bertossi12}.

\begin{example}\!\!\!\! \label{ex:paperML}(ex. \ref{ex:similarityComputation} cont.) We merge duplicate \ent{Paper} entities enforcing the MD:
{\small $\nit{Paper}$ $[\nit{pid}_1]\approx \nit{Paper}[\nit{pid}_2] \ \longrightarrow \ \nit{Paper}[\nit{Title},\nit{Year},
\nit{CID}, \nit{Keyword}] \doteq \nit{Paper}[\nit{Title},\nit{Year}, \nit{CID},$ $\nit{Keyword}]$}.\boxtheorem
\end{example}
 The portion, $\mc{P}^{\!M}$, of the general {\em LogiQL} program that represents  MD-based   merging contains rules as in {\bf 1.}-{\bf 4.} below:

\noindent {\bf 1.} \ The atoms of the form $R\mbox{-}\nit{Duplicate}$ mentioned above, and those representing the matching functions  (MFs)
$\match_{\!_A}$\!.

\vspace{2mm}
\noindent {\bf 2.} \ For an MD \ $R[r_1] \approx R[r_2] \ \longrightarrow \ R[\bar{r}_1] \doteq R[\bar{r}_2]$,  the rule:

\vspace{-5mm}
{\small\begin{eqnarray*}
&&R[r_1,\bar x_3]=\nit{bl}, \ R[r_2,\bar x_3]=\nit{bl} \ \longleftarrow \ R\mbox{-}\nit{Duplicate}(r_1,r_2), \ R[r_1, \bar x_1]=\nit{bl},\\ &&\hspace*{5.85cm}  R[r_2, \bar x_2]=\nit{bl}, \ \match(\bar x_1,\bar x_2)=\bar x_3,
\end{eqnarray*}}

\vspace{-6mm}
\noindent which creates two records (one of them can be purged afterwards) with different ids but all the other attribute values the same, and computed componentwise according to the
 MFs for $\match$. Here, $\bar{x}_1, \bar{x}_2, \bar{x}_3$ stand each for all attributes of relation $R$, except for the id and the block number (represented by $\nit{bl}$). (Block numbers play no role in merging.)

\vspace{2mm}
\noindent {\bf 3.} \ As for program $\mc{P}^{B}(D)$ given in Section \ref{MDBlocking}, rules specify the old versions of a record:

\vspace{-4mm}
{\small\begin{eqnarray*} && R\mbox{-}\nit{OldVersion}(r_1,\bar{x}_1) \ \leftarrow \ R[r_1, \bar{x}_1]=\nit{bl}, \ R[r_1, \bar{x}_2]=\nit{bl}, \ \bar{x}_1 \prec \bar{x}_2.
\end{eqnarray*}}

\vspace{-4mm}\noindent Here, $\bar{x}_1$ stands for all attributes other than the id and the block number; and on the RHS  $\bar{x}_1 \prec \bar{x}_2$ means componentwise comparison
of values according to the partial orders defined by the MFs.

\vspace{1mm}
\noindent {\bf 4.}~ Finally, rules to collect the latest version of each record, building the final resolved instance: \ \ \ \ {\small $R\mbox{-}\nit{ER}(r_1,\bar{x}_1) \;\;\leftarrow \;\;R[r_1,\bar{x}_1]= \nit{bl}, \ ! \ R\mbox{-}\nit{OldVersion}(r_1,\bar{x}_1)$.}

\vspace{2mm}
Notice that the derived tables $R\mbox{-}\nit{Duplicate}$ that appear in the LHSs of the MDs (or in the bodies of the corresponding rules) are all computed before (and kept fixed during) the enforcement of the merge-MDs. In particular, a duplicate relationship between any two records is not lost.
This has the effect of making the set of merging-MDs interaction-free, which results in a unique resolved instance.

\vspace{-4mm}
\section{Experimental Evaluation}\label{evaluation}

\vspace{-3mm}
We now show that our approach to ER can improve accuracy in comparison with standard blocking.  In addition to the MAS, we used datasets  from DBLP  and Cora Citation.

\begin{wrapfigure}[12]{r}{0.5\textwidth}
 \vspace{-35pt}
  \begin{center}
    \includegraphics[width=0.50\textwidth]{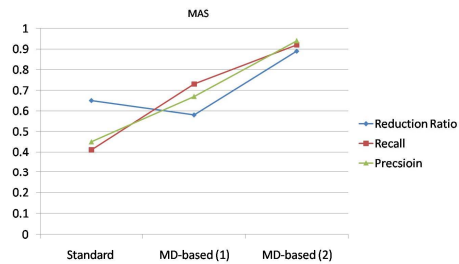}
  \end{center}
  \vspace{-25pt}
  \caption{The experiments (MAS)}\label{fig:plot12}
  \vspace{-25pt}
\end{wrapfigure}

 In order to emphasize the importance of semantic knowledge in blocking, we consider standard blocking and two different sets of MDs, (1) and (2), for MD-based collective blocking. Under (1), we define blocking-MDs for all the blocking keys used for standard blocking, but under (2) we have MDs for only some of the used blocking keys. In both cases, in addition to
 properly collective blocking MDs.

\vspace{5mm}
We use three measures for the comparisons of blocking techniques. One is {\em reduction ratio}, which is the
the ratio (minus $1$) of the number of candidate record-pairs over the initial number of records. The higher this value, the less candidate record-pairs are being generated, but the quality of the generated candidate  record pairs is not taken into account. We also use recall and precision measures. The former is the number of
true duplicate candidate record-pairs divided by the number of true
duplicate pairs, and  precision is the number
of true  candidate duplicate record-pairs divided by the total number of
candidate pairs \cite{Christen11}.

\begin{wrapfigure}{r}{0.5\textwidth}
  \vspace{-43pt}
  \begin{center}
    \includegraphics[width=0.50\textwidth]{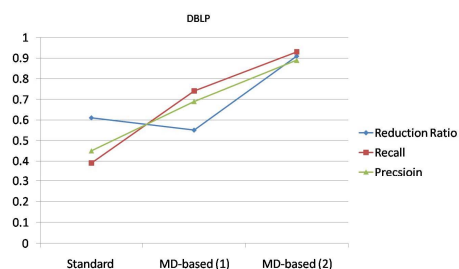}
  \end{center}
  \vspace{-24pt}
  \caption{The experiments (DBLP)}\label{fig:plot11}
  \vspace{-25pt}
\end{wrapfigure}

\begin{wrapfigure}[10]{r}{0.5\textwidth}
  \vspace{-35pt}
  \begin{center}
    \includegraphics[width=0.50\textwidth]{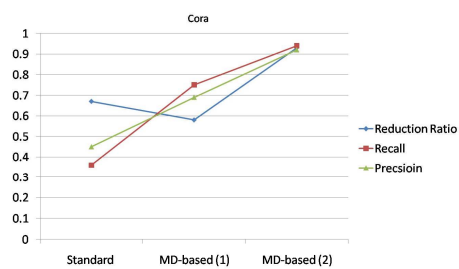}
  \end{center}
  \vspace{-28pt}
  \caption{The experiments (Cora)}\label{fig:plot13}
  \vspace{-27pt}
\end{wrapfigure}

Figures \ref{fig:plot12}, \ref{fig:plot11} and \ref{fig:plot13}  show the comparative performance of {\em ERBlox}. They show that standard blocking has higher reduction ratio than MD-based collective blocking version ($1$). This means that less candidate record-pairs are being generated by standard blocking. However, the precision  and recall of  MD-based blocking version ($1$) are higher than standard blocking, meaning that MD-based blocking version ($1$) can lead to improved  ER results at the
cost of larger blocks, and thus more candidate record pairs that need to be compared.

 In blocking, this is a common tradeoff that needs to be considered. On the one hand, having a large number of smaller blocks will result in fewer candidate record-pairs that will be generated, probably increasing the number of true duplicate record-pairs that are missed. On the other hand, blocking techniques that result in larger blocks generate a higher number of candidate record-pairs that will likely cover more true duplicate pairs, at the cost of having to compare more candidate pairs \cite{Christen11}. The experiments are all done before MD-based merging.

Interestingly, MD-based blocking version ($2$) has higher reduction ratio, recall, and precision than standard blocking. This emphasizes the importance of MDs supporting collective blocking, and shows that blocking based on string similarity alone fails to capture the relationships that naturally hold in the data.

\vspace{8mm}As expected, the experiments show that different sets of MDs for MD-based collective blocking have different impact on reduction ratio, so as standard blocking depends on the
 choice of blocking keys.  However, the quality of MD-based collective blocking, in its two versions, dominates standard blocking for the three datasets.

\ignore{\section{Related Work}\label{RelWork}
Some previous work falls into the bucket of declarative data cleaning. We note however that the functionality that can be specified declaratively often differs significantly from one work to another, and from ours.

In this direction, some recent work has considered collective deduplication in the presence of constraints. For instance, \cite{Arasu09} proposes a declarative framework for collective deduplication of entities in the presence of constraints. Constraints are represented by a simple declarative Datalog style language. However, this framework does not merge the duplicate records into a single representation for ER. In our work, we propose MD-based merging component to do so.

Previous work has considered the use of Markov Logic Networks and other related probabilistic models to do collective
deduplication with constraints. A representative work in this direction is \cite{Singla06}. This class of work provides clear semantics.
The output is (conceptually) a space of possible deduplications in contrast to a single deduplication that we generate in our work.

Another line of research is blocking techniques. Various blocking techniques have been proposed. For a comprehensive survey on blocking see \cite{Christen11}. However, most prior approaches to blocking
are inflexible for at least one of two reasons: (1) They
allow blocking of only a single entity type in isolation, (2) They ignore valuable domain or semantic knowledge that can be used for blocking.

In this respect, the only work that briefly applies semantic knowledge  for blocking is \cite{Rastgoi11}. The paper proposes a framework for large-scale collective entity matching that applies similarity of blocking keys and relational closeness for blocking of entities.  However,  the paper does not mention how the relational closeness are represented and how they are captured.

Iterative blocking has been proposed in \cite{Garcia-Molina09}, which the ER results of blocks are reflected to subsequently
processed blocks. Blocks are now iteratively processed until no block contains any more matching records. In this way, the iterative blocking reduces
false negatives (i.e., improve recall). However, the iterative blocking does not apply semantic knowledge.

We use MDs to express semantic knowledge in our work. \cite{Song09} studies the problem of discovering matching dependencies (MDs) for a given
database instance. First, they define the measures, support
and confidence, for evaluating the utility of MDs in the given
database instance. Then, they study the discovery of MDs with certain
utility requirements of support and confidence.}

\ignore{
Efficient algorithms to
support the framework. Our algorithms have precise theoretical
guarantees for a large subclass of our framework. We show,
using a prototype implementation, that our algorithms scale to
very large datasetsis used for identifying groups of tuples
that could be merged. However, they do not do the merging
(a main contribution in our approach) or use MDs

Recently,

Some recent work has considered collective deduplication
in the presence of constraints

Clustering is a well-studied problem and has applications
well beyond data cleaning. The previous work is that is most
related to this paper is [26], which contains new algorithms for
correlation clustering. Our main algorithms are generalizations
of one of the algorithms in [26]. The correlation clustering
problem was first proposed and studied by Bansal et al. [24].

There have been various similarity measures available
e.g., edit distance [17], token-based n-gram [13], and
character-based qgram [29].

\subsection{Blocking}
The most important ER blocking methods are summarized in \cite{Christen11}.

It is possible to perform MD-based blocking iteratively as proposed in \cite{Garcia-Molina09} where the ER results of blocks are reflected to subsequently processed blocks. Blocks are now iteratively processed until no block contains any more matching records.

\subsection{MDs}
\red{Shaoxu Song, Lei Chen, "Efficient discovery of similarity constraints for matching dependencies", Data Knowledge Engineering, Elsevier,   S. Song, L. Chen, Discovering matching dependencies, CIKM}

\red{A key
feature of our language is that it allows users to specify both
hard and soft constraints}

\comzb{should we change the attribute values in MD-based blocking using MDs? if we change them we also enrich the data for the next phase which is detection identification. none of the existing blocking techniques do the above point}}

\vspace{-0.4cm}

\section{Conclusions}\label{conclude}

\vspace{-3mm}
We have shown that matching dependencies, a new class of data quality/cleaning semantic constraints in databases, can be profitably integrated with traditional
ML-methods, in our case for entity resolution. They play a role not only in the intended goal of merging duplicate representations, but also in the
record blocking process that precedes the learning task.  At that stage they allow to declaratively capture semantic information that can be used to enrich the blocking activity.
MDs declaration and enforcement, data processing in general, and machine learning  can all  be integrated using the {\em LogiQL} language.

\vspace{1mm}\noindent {\small {\bf Acknowledgments:} \ Part of this research was funded by an NSERC Discovery grant and
the NSERC Strategic Network on Business Intelligence (BIN). \ Z. Bahmani and L. Bertossi are very much grateful for the
support from LogicBlox during their internship and sabbatical visit. }

\ignore{\red{we discuss two practical extensions to our
work: adding weights to MDs and support for reference tables.}

\red{iterative blocking}

\red{Use  prior frameworks such as Markov
Logic Networks [12] or Probabilistic Relational Models [13] for blocking}

\red{The experiments show promising
results for recall, and most importantly its significant increasing
when rules are added.}

\red{the collective EM techniques suffer from a fundamental
problem: their poor scalability. The high cost of probabilistic
inference over large EM graphs renders these methods computationally
infeasible for large data sets.}

When variables $z_i, z_j$ in (\ref{eq:newMDs}) correspond to attributes of relations in $\psi$, not $\nit{Bl\#}$, the defined blocking MDs is used for enriching the attributes values that are considered to be a part of blocking keys. Moreover, enforcing these MDs leads to more informative data and more complete piece of information. This is significantly important since the MD-based blocking not only reduces the large amount of potential record pair comparisons, but also enriches information for record duplicate detection. Thus, the MD-based blocking can lead to improved matching or deduplication results for the second component of ERBlox.}

\ignore{Appendix
\newpage

\appendix
\section{Appendix}

\subsection{Merge-MDs are interaction-free}

\begin{example} Assume that we have the following relation \ent{RecDuplicate} which contains positive cases from the classifier. Here, we have transitive positive cases, and we need to merge them. Consider we have the original database instance $D$. If we start merging from $\langle r_1, r_2\rangle$,  then we have $D_1$:

{\footnotesize
\begin{center}
\begin{tabular}{c|c|c|}\hline
$\nit{RecDuplicate}$&$\nit{rid}_1$ & $\nit{rid}_2$\\ \hline
$$&$r_1$&$r_2$\\
$$&$r_2$&$r_3$ \\\cline{2-3}
\end{tabular}\hspace*{0.5cm}
\begin{tabular}{c|c|c|c|}\hline
$R(D)$&$A$ & $B$&$C$\\ \hline
$r_1$&$a_1$&$b_1$& $c_1$\\
$r_2$&$a_2$&$b_2$&$c_2$ \\
$r_3$&$a_3$&$b_3$&$c_3$ \\\cline{2-4}
\end{tabular}\hspace*{0.5cm}
\begin{tabular}{c|c|c|c|}\hline
$R(D_1)$&$A$ & $B$&$C$\\ \hline
$r_1$&$\langle a_1,a_2\rangle$&$\langle b_1,b_2\rangle$& $\langle c_1,c_2\rangle$\\
$r_2$&$\langle a_1,a_2\rangle$&$\langle b_1,b_2\rangle$&$\langle c_1,c_2\rangle$ \\
$r_3$&$a_3$&$b_3$&$c_3$ \\\cline{2-4}
\end{tabular}
\end{center}}
\noindent Next, we merge $\langle r_2,r_3\rangle$ which results in instance $D_2$. Again we need to merge $\langle r_1, r_2\rangle$ because they are a positive case according to the classifier. Then the final ER result would be instance $D_3$ :

{\footnotesize
\begin{center}
\begin{tabular}{c|c|c|c|}\hline
$R(D_2)$&$A$ & $B$&$C$\\ \hline
$r_1$&$\langle a_1,a_2\rangle$&$\langle b_1,b_2\rangle$& $\langle c_1,c_2\rangle$\\
$r_2$&$\langle a_1,a_2, a_3\rangle$&$\langle b_1,b_2,b_3\rangle$&$\langle c_1,c_2,c_3\rangle$ \\
$r_3$&$\langle a_1,a_2, a_3\rangle$&$\langle b_1,b_2,b_3\rangle$&$\langle c_1,c_2,c_3\rangle$ \\\cline{2-4}
\end{tabular}\hspace*{0.3cm}
\begin{tabular}{c|c|c|c|}\hline
$\nit{ER}(D_3)$&$A$ & $B$&$C$\\ \hline
$r_1$&$\langle a_1,a_2, a_3\rangle$&$\langle b_1,b_2,b_3\rangle$& $\langle c_1,c_2,c_3\rangle$\\
$r_2$&$\langle a_1,a_2, a_3\rangle$&$\langle b_1,b_2,b_3\rangle$&$\langle c_1,c_2,c_3\rangle$ \\
$r_3$&$\langle a_1,a_2, a_3\rangle$&$\langle b_1,b_2,b_3\rangle$&$\langle c_1,c_2,c_3\rangle$ \\\cline{2-4}
\end{tabular}
\end{center}}
\noindent Even if we start from merging $\langle r_2, r_3\rangle$, then we would have the same result because the content of  predicate \nit{RecDuplicate} is not changed. We do not need to compute the transitivity of the positive cases before merging. The merging-MDs take care of that.\boxtheorem
\end{example}

\subsection{Whatever this is ...}

\begin{verbatim}
/*******************Character in each position***********/
fname_char_at[x,i]= c <- (x),_fname[x]=fname,num=string:
length[fname],dom(i), i<num, string:substring[fname,i,1]
=c, string:length[c]>0, c!=".",  c!="'",  c!="-",  c!=" ".
lname_char_at[x,i]= c <- (x),_lname[x]=lname,num=string:
length[lname],dom(i),i<num, string:substring[lname,i,1]
=c, string:length[c]>0, c!=".",  c!="'",  c!="-",  c!=" ".
aff_char_at[x,i]= c <- (x),_aff[x]=aff,num=string:
length[aff],dom(i), i<num, string:substring[aff,i,1]=
c, string:length[c]>0, c!=".",  c!="'",  c!="-",  c!=" ".
fname_leng[x]=m <- agg<< m=count()>>fname_char_at[x,_]=_.
lname_leng[x]=m <- agg<< m=count()>>lname_char_at[x,_]=_.
aff_leng[x]=m <- agg<< m=count()>>aff_char_at[x,_]=_.

/*******************Jaro-winkler************/
lname_match(id1, id2, i1, i2, c) <-(id1),(id2),
rblock(id1, id2), num1=lname_leng[id1],num2=lname_leng
[id2], num2<num1, p1=(num1/2),lname_char_at[id2,i2]=c,
i1-i2<=p1-1,lname_char_at[id1,i1]=c.
lname_match(id1, id2, i1, i2, c) <-(id1),(id2),
rblock(id1, id2), num1=lname_leng[id1],num2=lname_leng
[id2], num2<num1, p1=(num1/2),lname_char_at[id2,i2]=c,
 i2-i1<=p1-1,lname_char_at[id1,i1]=c.
lname_match(id1, id2, i1, i2, c) <-(id1),(id2), rblock
(id1, id2), num1=lname_leng[id1],num2=lname_leng[id2],
num1<num2, p1=(num1/2),lname_char_at[id2,i2]=c, i1-i2
<=p1-1,lname_char_at[id1,i1]=c.
lname_match(id1, id2, i1, i2, c) <-(id1),(id2),
rblock(id1, id2), num1=lname_leng[id1],num2=lname_leng
[id2], num1<num2, p1=(num1/2),lname_char_at[id2,i2]=c,
 i2-i1<=p1-1,lname_char_at[id1,i1]=c.
lname_match(id1, id2, i1, i2, c) <-(id1),(id2), rblock
(id1, id2), num1=lname_leng[id1],num2=lname_leng[id2],
 num2=num1, p1=(num1/2),lname_char_at[id2,i2]=c, i1-i2
 <=p1-1,lname_char_at[id1,i1]=c.
lname_match(id1, id2, i1, i2, c) <-(id1),(id2), rblock
(id1, id2), num1=lname_leng[id1],num2=lname_leng[id2],
 num2=num1, p1=(num1/2),lname_char_at[id2,i2]=c, i2-i1
 <=p1-1,lname_char_at[id1,i1]=c.
lname_trans_tmp2(id1,id2,i11,i21,i12,i22) <-
lname_match(id1,id2,i11,i21,c1),lname_match(id1,id2,
i12,i22,c2),i11< i12, i21>i22.
lname_trans_count(id1,id2,0) <- rblock(id1,id2),
!lname_trans_tmp2(id1,id2,n1,m1,n2,m2).
lname_trans_count(id1,id2,t)<-lname_trans_count_tmp
[id1,id2]=t.
lname_trans_count_tmp[id1,id2]=t <- agg<< t=count()
>>lname_trans_tmp2(id1,id2,_,_,_,_).
lname_match_count[id1,id2]=m <- agg << m=count()
>>lname_match(id1,id2,_,_,_).
lname_jaro_tmp(id1,id2, n) <- num1=lname_leng[id1]
, num2=lname_leng[id2], (id1),(id2),lname_trans_count
(id1,id2,t),lname_match_count[id1,id2]=m, m>0,
n= float64:round[m]/float64:round[num1]+float64:round[m]
/float64:round[num2]+(float64:round[m]-float64:
round[t/2])/float64:round[m].
lname_jaro(id1,id2, num) <-lname_jaro_tmp(id1,id2,n)
, num=n/3.0.
lname_wv_part(id1,id2,0) <- lname_match_count[id1,
id2]=m, m=0.
lname_jaro(i1,i2,num) ->(i1),(i2), float[64](num).
lname_winkler_tmp(i1,i2, num) -> (i1),(i2),
float[64](num).
lname_wv_part(id1,id2, num2) <- lname_jaro(id1,id2,
 num), num2=num+(2.0*0.1*(1-num)).
/*******************Cbigram************/
aff_bichar_at[x,i]= c <- (x),_aff[x]=aff,num=string:
length[aff],dom(i), i<num, string:substring[aff,i,2]=c,
 string:length[c]>0.
aff_chars(x,c) <- aff_char_at[x,i]=c.
aff_uni_chars(i1,i2,c) -> (i1),(i2),string(c).
aff_uni_chars(i1,i2,c) <- aff_chars(i1,c), rblock
(i1,i2).
aff_uni_chars(i1,i2,c) <- aff_chars(i2,c), rblock
(i1,i2).
aff_uni_count[i1,i2]=num -> (i1),(i2),int[64](num).
aff_uni_count[i1,i2]=n <- agg << n=count() >>
aff_uni_chars(i1,i2,c).
aff_com_count[i1,i2] =num -> (i1),(i2),int[64](num).
aff_com_count[i1,i2]=n <- agg<< n=count()>> aff_
chars(i1,c),aff_chars(i2,c),rblock(i1, i2).
aff_wv_part(i1,i2,pct) <- aff_com_count[i1,i2]=n1,
 aff_uni_count[i1,i2]=n2, pct = float64:divide[n1, n2].
/*******************weight vectors************/
wv(id1, id2,fnamenum,lnamenum,affnum) <-fname_wv_part
(i1, i2,fnamenum),lname_wv_part(i1, i2,lnamenum),
aff_wv_part(i1, i2,affnum),(i1),(i2),:id(i1:id1),
:id(i2:id2),id1<id2.
\end{verbatim}

}
\end{document}